\documentclass[sn-basic]{sn-jnl}

\jyear{2022}%

\theoremstyle{thmstyleone}%
\theoremstyle{thmstyletwo}%

\theoremstyle{thmstylethree}%

\begin{document}

\title[Hybrid Probabilistic-Snowball Sampling]{Hybrid Probabilistic-Snowball Sampling Design}


\author[1]{\fnm{Giulio Giacomo} \sur{Cantone}}\email{giulio.cantone@phd.unict.it}

\author*[2]{\fnm{Venera} \sur{Tomaselli}}\email{venera.tomaselli@unict.it}

\affil[1]{\orgdiv{Department of Physics and Astronomy ``E. Majorana"}, \orgname{University of Catania}, \orgaddress{\street{S. Sofia, 64}, \city{Catania}, \postcode{95123}, \state{Italy} \country{IT}}}

\affil*[2]{\orgdiv{Department of Political and Social Sciences}, \orgname{University of Catania}, \orgaddress{\street{Vittorio Emanuele II, 8}, \city{Catania}, \postcode{95131}, \state{Italy} \country{IT}}}


\abstract{Snowball sampling is the common name for sampling designs on human populations where respondents are requested to share the questionnaire among their social ties. With some exceptions, estimates from snowball samplings are considered biased. However, the magnitude of the bias is influenced by a combination of elements of the sampling design and features of the target population. Hybrid Probabilistic-Snowball Sampling Designs (HPSSD) aims to reduce the main source of bias in the snowball sample through randomly oversampling the first stage 0 of the snowball.

To check the behaviour of HPSSD for applications, we developed an algorithm that, by grafting the edges of a stochastic blockmodel into a graph of cliques, simulates an assortative network of tobacco smokers. Different outcomes of the HPSSD operations are simulated, too.

Inference on 8,000 runs of the simulation leads to think that HPSSD does not improve reliability of samples that are already representative. But if homophily in the population is sufficiently low, even the unadjusted sample mean of HPSSD has a slightly better performance than a random, but undersized, sampling.

De-biasing the estimates of HPSSD shows improvement in the performance, so an adjusted HPSSD estimator is a desirable development.}


\maketitle

\section{Introduction}\label{Intro}

In population studies, designs of sampling procedures involving randomisation in the process of drawing the respondents are recognised as probabilistic designs. Probabilistic designs, if properly implemented, reduce confounding and colliding effects in observed outcomes. In this case, the sample statistics are assumed to be unbiased estimators of features in the whole population. Probabilistic designs are operationally expensive but reliable for surveys aimed at mapping novel phenomena in social sciences. So, it is correct to consider them the `gold standard' of population studies.

However, as remarked by Groves  (\citeyear{groves_three_2011}), operational costs (e.g., costs in working hours) still represent a serious burden to conduct high quality research for replication studies, monitoring, \textit{censi}, etc.

This is even more truthful in the Internet Era. Non-responses (attrition) are the Achilles' Heel of probabilistic designs, because it increase the operational cost of surveys. There are evidences that attrition rates has increased over years \citep{de_heer_trends_2002,bethlehem_solving_2016,williams_trends_2018}. For example, the fact that mobile phones have been adopted worldwide increased, not decreased, the operational costs of traditional telephonic surveys \citep{vicente_initial_2017}. With exception of some sensible survey (e.g. on political opinions), it is assumed that non-responses are missing at random: missing data is uncorrelated with observed outcomes of the survey. This assumptions does not hold always \citep{weidmann_missing_2021}, so a raise in attrition rates would not only increase operational costs, but also bias the results.

On parallel, the adoption of non-probabilistic sampling designs has grown in social sciences\citep{lehdonvirta_social_2021}. Non-probabilistic designs are not justified through Probability Theory alone. Sometimes contextual features of the research or robust prior knowledge can justify alone the adoption of a non-probabilistic design\footnote{The theory behind the adoption of a non-probabilistic sampling design should be contextual to a phenomenon that is already well studied with other tools. The political poll conducted among Xbox gaming players by Wang et al (\citeyear{wang_forecasting_2015}) and the adopted weighting scheme is a notorious case of a non-probabilistic (`non-representative') design that led into an accurate forecast.}.

\begin{figure}[h]
\label{fig:forest}
\centering
\includegraphics[width=0.7\textwidth]{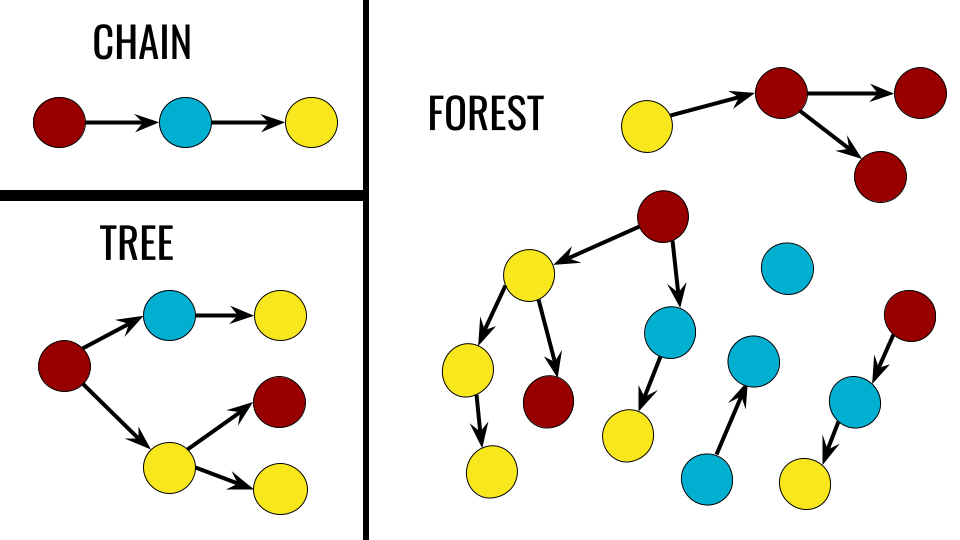}
\caption{The tree is a spatial chain that can fork itself into different sub-chains. Colours represent discrete states of one attribute. A forest is a graph made of trees. The apex node of the tree is called seed. In a forest, a seed alone without children nodes can represent a sub-graph of the forest. The $t$-stage of a node is the number of steps necessary to reach that node from the seed.}
\end{figure}

However, there is specific non-probabilistic survey design that is, more often than not, problematic at its core: when a person or few people ask to their own social ties to fill a survey tool (i.e., a questionnaire). The contacted people can also be encouraged to share the survey tool among their own social ties. This process of `responding-sharing-responding' can be modeled as a Galton-Watson branching process tree, that is a conceptual expansion of the more common concept of discrete Markov Chain \citep{rohe_critical_2019}. An ensemble of trees will be then called `a forest' (Figure \ref{fig:forest}).

From this parallel, one can recognize the pitfall of this design: for any non-trivial correlation between stages (i.e., correlation between the state of the recruiter and the recruited), the final sample will be dependent on the random outcomes in early stages. With some exceptions \citep{spreen_rare_1992}, sample statistics will be biased.

This process is sometimes called, maybe improperly \citep{goodman_comment_2011}, `snowball sampling'. In the practice, it happens that the first stage of responses is not even randomly drawn, and that further responses are collected until a sample size deemed sufficient is reached. If this is the case, it is hard to image that the final sample could it be representative of the target population.

Correlations between stages of a recruitment forest happen because:

\begin{itemize}
    \item The underlying population is assortative: nodes have a general preferential attachment for connecting to nodes with some characteristics and not others \citep{mcpherson_birds_2001, cantwell_friendship_2021, evtushenko_paradox_2021}.
    \item The mechanism generating the forest is assortative: by design or just by individual preferences, something is biasing the specific collection of respondents \citep{crawford_identification_2018}.
\end{itemize}
When assortative mechanics bring out connections among nodes with similar features, it is said that there is homophily among the nodes in the graph.

Since the final outcome depends on the early stages, in our opinion, the practical problem is better framed as a problem of sample size and sampling design of the fraction of sample units collected at the stage 0, the seeds.

In this paper, a computational study is carried out to enquire if sufficient conditions exist for allowing an estimation better or equally good than probabilistic sampling designs, but with reduced operational costs. This proposal is called Hybrid Probabilistic-Snowball Sampling Design (HPSSD).

To assert this result, we developed a computational simulation (Section \ref{Meth}) in two parts:

\begin{enumerate}
    \item an algorithm that simulates a network where the population of nodes can be homophile both regarding a binary variable and the number of connections (degree)
    \item another algorithm that simulates a HPSSD in the artificial population.
\end{enumerate}

This procedure is iterated $8,000$ times (Monte Carlo), each mutually independent and initiated with random hyperparameters. As a reference case, a network of people is modeled with around a quarter of nodes as tobacco smokers (Section \ref{Theo}).

Inference is performed on summary statistics over the $8,000$ independent runs. Results (Section \ref{Resu}) induces to think that even small homophily would make the HPSSD less reliable than the costwise alternative random sampling. However, even a coarse technique to reduce bias in the estimator would make HPSSD consistently performing better than the gold standard. Developments, heuristics, generalizability, and other limitations for the study are discussed in Section \ref{Disc}.

\section{Theoretical Background}\label{Theo}

In Section 1, snowball sampling has been presented as a method employed by qualitative researchers, decoupled from the problem of estimation \citep{biernacki_snowball_1981}. However, as remarked by the first proponent of a `snowball' sampling, Leo Goodman (\citeyear{goodman_comment_2011}), this description stems from a misconception. Goodman's model \citep{goodman_snowball_1961} was originally aimed to treat analytically the methodology of data collection pioneered by the team of sociologist James Coleman (\citeyear{coleman_relational_1958}). Originally, the first stage 0 of snowball sampling was supposed to be a randomised drawing of a small number of units from the target population and not any available set of eligible participants \citep{granovetter_network_1976, frank_sampling_1978, rapoport_probabilistic_1979}.

In the Goodman's model, after the first draw, each sampled unit is asked to recruit a fixed $m$ of other respondents within the target population:
\begin{equation} \label{Eq:goodman}
n_t = n_{t-1} \cdot m; n_0 = n_{seeds}
\end{equation}
$m$ being fixed implies in Eq. \ref{Eq:goodman} that zero attrition is expected in the recruitment process. In this case the proprieties of the tree can be derived, with only minor adjustments, from the proprieties of the Markov Chain \citep{lu_sensitivity_2012,rohe_critical_2019}.

The tree model corrected with $r$ attrition
\begin{equation} \label{Eq:goodmanr}
n_t \sim n_{t-1} \cdot m (1 - r); n_0 \sim n_{seeds} \cdot (1 - r)
\end{equation}
is not anymore an exact model, but a stochastic one. It follows that knowing the proportion of non-responses in the sample $r$, then it is possible to model the distribution of $r_i$ as if each unit has an individual attrition $r_i$.

Network sampling has been revamped by the works of Frank and Snijders (\citeyear{frank_estimating_1994}) and Heckathorn (\citeyear{heckathorn_respondent-driven_1997}), under a new name: Respondent-Driven Sampling (RDS). RDS has a different axiomatization than snowball sampling:

\begin{itemize}
    \item RDS has always a finite `target' population, that is also always a network.
    \item $m_i$ is allowed to variate across sample units $i$
    \item each respondent unit has to self-report the $k_i$ number of its social ties within the target population
    \item RDS lacks the assumption that seeds are randomly drawn. While for some target population, this is quite convenient, it has been demonstrated to be a flaw more than a strength. For example, Khabbazian et al (\citeyear{khabbazian_novel_2017}) proposed an alternative, Anti-Clustering RDS, expressively to avoid the issue of intra-cluster branching of the respondents.
\end{itemize}

An interesting methodological introduction of RDS is the explicit implementation of link-tracing of the links within the forest of respondents. This implementation has been discussed even between introduction of Internet in households \citep{spreen_rare_1992}.

RDS theory and RDS estimators \citep{volz_probability_2008} have been object of (self) criticism, for example regarding: the asymptotic proprieties for the sample size \citep{verdery_new_2017}; subjectivity in estimation of $k_i$ \citep{lu_sensitivity_2012}; and unavoidable biases in variance estimators  \citep{goel_assessing_2010, ott_unequal_2016, verdery_new_2017}.

Crawdford, Aronow, Zheng, and Li (\citeyear{crawford_identification_2018}), partially basing on the previous work of Gile and Handcock (\citeyear{gile_7_2010}) and Tomas and Gile (\citeyear{tomas_effect_2011}), pointed out that in the case of highly attribute-assortative forests, $k_i$ is not a sufficient information for unbiased inference in a network. This problem is exacerbated if non-responses\footnote{The problem of attrition is the characteristic issue of population studies as an empirical social science: this problem is virtually absent in applications of network sampling designs originated outside social sciences and only then applied for inferences on human populations. For example, network-crawling techniques applied on Social Media \citep{leskovec_sampling_2006, gjoka_walking_2010}, although considered very successful for demographic inference, do not assume any agency in the nodes, i.e. nodes cannot refute to be surveyed.} are biased by underlying attribute-assortativity, too \citep{smith_network_2017}.

The standard measure of assortativity is the Bravais-Pearson's linear correlation. This standard holds across different data formats, since for binary attributes the correlation's coefficient is reduced to Pearson's $\phi$ for pairs of connected nodes with same or different values \citep{mcpherson_birds_2001, newman_networks_2010}, with only minor differences between directed and undirected networks. While other measures of correlation have been proposed \citep{noldus_assortativity_2015}, the assumption of linearity of the correlations is paradigmatic.

In most application, $\phi$ is correctly measured on the whole population or very large samples\footnote{Inference of characteristics of a network from a sample is an advanced task, because a summary statistic of the graph cannot be decomposed as a linear function of sample space of the population of the nodes. For example, it holds:
\begin{equation*}
l[\phi(\textbf{a},\textbf{b}),\phi(\textbf{c},\textbf{d})] \neq \phi [l(\textbf{a},\textbf{c}), l(\textbf{b},\textbf{d})]
\end{equation*}
for any $l$ linear function (e.g., the average). So to infer features of the network (not of the nodes) the sampling design is aimed to sample not a collection of nodes, but a collection of subgraphs that are representative projections of the whole graph \citep{leskovec_sampling_2006, ahmed_network_2013, crane_probabilistic_2018}. Interestingly, in this case random sampling is not `gold standard' and it is inefficient to reach a representative sample of subgraphs of the whole network \citep{zhang_graph_2017}.}.

\subsection{Hybrid Design}

In Hybrid Probabilistic-Snowball Sampling Design (HPSSD) a fraction of respondents is recruited with a probabilistic procedure, and a subsequent fraction is recruited by the first fraction. Differently from Goodman's model, $m_i$ is not fixed: seeds are required to spread the survey tool as much as possible among their social ties.

What is the implication of the formulation ``as much as possible" in terms of statistical distribution of $m_i$-recruitments within the tree? Social networks have a tendency to generate scale-free distribution of $k_i$-degrees \citep{fortunato_scale-free_2006,barabasi_scale-free_2009}, hence in a scenario where $m_i$ is correlated with $k_i$ it may happen that many chained respondents share a small number of common seeds, while the other seeds are `infertile'.

However, we would argue that the case for a exceptionally large tree dominating a forest, while possible, is not the typical scenario of a HPSSD. We expect seeds to just share the survey tool in more private and family-oriented social media, since these are possibly the most efficient platforms to informally recruit social ties, compared to alternatives \citep{baltar_social_2012, brickman_bhutta_not_2012,herbell_facebook_2018,lindsay_faith_2021}. This would be a case when even nodes with high $k_i$ would recruit a modest $m_i$.

\subsection{Application on tobacco smoking}

An interesting antecedent has been designed in Etter and Perneger (\citeyear{etter_distributions_1997}) in the context of sampling tobacco smokers in Geneva, Switzerland, in 1999. Authors randomly sampled $3300$ inhabitants through a register of email addresses. These have been equipped with a virtual coupon and then asked
\begin{itemize}
    \item if smokers or ex-smokers (target): to fill a questionnaire and to send it back to researchers through an online procedure, with the coupon number. Coupons returned in this stage are the primary component of the sample.
    \item if not target: to ask to any known person within target to fill the questionnaire, and to send it back with the coupon number of the seed. Coupons returned in this stage are the secondary component of the sample.
\end{itemize}
$578$ questionnaires have been returned in the primary component of the sample. With an estimated smoking prevalence of $.32$, the estimated attrition is $1 - \frac{578}{.32 \cdot (3300)} \sim .45$.

The respondents in the secondary component were $566$. This is significantly lower than the expected value of $.68 \cdot (3300) \cdot (1-.45)^2 = 678$ (see, Eq. \ref{Eq:goodmanr}), meaning that at least one of the two average attrition rates ($\bar{r_i}$) in the two sample components is higher than expected.

Nevertheless, authors report that not only the two components showed only minor statistically significant differences (in particular, a small gender prevalence in the secondary component), but also that the estimates on the combined hybrid sample were not statistically different from a previous representative benchmark \citep{etter_distributions_1997}. Authors explained the performance of the union of the two components through the overall sample size of the primary component of the sample (seeds), that is randomly drawn.

For a population of smokers (in Geneva, 1999) that could not exceed $40.000$ people, a random sample of $578$ is associated to a maximum margin of error of $\frac{.98}{\sqrt{578}} \cdot 100 = 4.07\%$\footnote{Here is applied the standard Laplace's formula for the margin of error $1.96 \cdot \sqrt{\frac{\sigma^2}{n}} \cdot 100$ that maximises the expected entropy assuming that the target variables are uniformly split.} with Confidence Level $95\%$. While this number is too high to consider $n = 578$ a truly `representative' sample, is not excessively high. 
Why even the secondary component alone performed so well for the estimation? A simple explanation for it is that, given a fraction of random seeds, even not representative of a population, but close to an acceptable margin of error, then the snowball sample performs as a representative sample. This is the hypothesis of the present study on HPSSD.

However, other explanatory elements could occur. For example, correlation between attrition ${r_i}$ and smoking can be only weak \citep{mccoy_attrition_2009,powers_impact_2010,zethof_attrition_2016}, and very likely correlated through a mediation effect, e.g. level of scientific education \citep{siddiqui_factors_1996,cunradi_survey_2005,young_attrition_2006,haring_extended_2009,mcdonald_implications_2017}.
A discussion on assortativity among smokers requires a more complex analysis. There are strong evidences that family is the main driver of smoking status \citep{otten_parental_2007}. Smokers tend to start families together \citep{clark_dont_2006, agrawal_assortative_2006, malagon_assortativity_2017} and their smoking status is then culturally inherited by their children \citep{charlton_children_1996,bricker_prospective_2006,leonardi-bee_exposure_2011}. This is a case of complex social contagion \citep{centola_how_2018} in the sense of mutually reinforced (or, looped) causality. In another example of social influence within households: when married people decide to quit smoking, both of them, as individuals, are less successful in it if the other partner ceases to quit smoking\citep{waldron_family_1989,christakis_collective_2008,blonde_cohabitation_2022}.

The second driver of assortativity in smokers regards how and why smokers bond with family-unrelated peers. When smoking has been seen as a harmless element of fashion, smokers were the most central individuals in social networks, but after smoking was associated with diseases, the quota of smokers plummeted, and the smokers were clustered into peripheral areas of the social network \citep{christakis_collective_2008, philip_relationship_2022}. There is relevant debate if a process of transitions from never-smoker to smoker to ex-smoker can be called a social `contagion' (`influence'). Aral, Muchnik, and Sundararajan (\citeyear{aral_distinguishing_2009}), and Shalizi and Thomas (\citeyear{shalizi_homophily_2011}) have been opposed to these definitions because social influence usually is not identified without the mediation effect of pre-contagion assortativity, that is the association between being linked as social ties and the factors of risk of falling into smoking status. In this sense, it would be a common environment driver and not a direct influence \citep{go_social_2012, cheadle_differential_2013}.

Hence, we propose a model of assortative networks where connections are not driven by the target variable (smoking), but by its risk score, that is a summary number in the unit interval that measures the likelihood of the expected outcome (smoker/non-smoker).

Here the risk score can resemble a propensity score \citep{austin_introduction_2011}, but it is only a numerical abstraction useful for generating a sophisticated artificial network of smokers, instead. Nodes would be degree-assortative and smoking-assortative as a epiphenomenon of the homophily in their propensity to be smoker.

In detail, the individual risk score is a compact summary of the joint effect of a multivariate distribution of predictive factors of smoking. So, if it true that there is a common inheritance of these factors within the household, this can be represented as a clique (that could be even a single person) sharing a common coefficient of the risk score. Then the individual risk score could be represented as a function of the coefficient fixed within the clique and a random coefficient. From this model to represent determination of the propensity to smoke, stem the random network generating model that we called ``cliques-and-blocks", presented in Section \ref{PopModel}.

\section{Methods}\label{Meth}

\subsection{Population Generation Model: cliques-and-blocks} \label{PopModel}

The goal for the algorithm is to generate a network of smokers and not smokers (binary attribute $y$). The binary attribute follows a Bernoulli distribution. The parameter of the Bernoulli is a risk score that is a mixed value from a $\alpha$-coefficient fixed within clique and a second random $\beta$-coefficient. We want the network to be assortative in regards of the $y$ binary attribute.

To achieve this goal, the algorithm has to:
\begin{enumerate}
    \item draw a set of $\omega$-cliques with a fixed $\alpha$ within clique;
    \item randomly assign $i$-nodes into the cliques and then assign $\beta$ to $i$;
    \item compute a mixing function:
\begin{equation}\label{eq:p}
    e(i) = \alpha_{\omega} \cdot w + \beta_i \cdot (1-w)
\end{equation}
where $\beta_i$ and $\alpha_{\omega}$ are values drawn from the same $D$ distribution on the unit interval, and $w$ is a weight parameter, fixed for the whole population.
\end{enumerate}
A tabular representation of this scheme is provided in Table \ref{tab:ei}.
\begin{table}[h]
\begin{center}
\begin{minipage}{200pt}
\caption{Example of a table of attributes for the $i$ nodes.}\label{tab:ei}%
\begin{tabular}{@{}llllll@{}}
\toprule
$i$ & $\omega$-clique & $\beta_i$ & $\alpha_{\omega}$ & $e(i)$ & $y_i$ \\
\midrule
1 & 1 & $\beta_1$ & $\alpha_1$ & $e \mid (\beta_1,\alpha_1)$ & $y_1 \mid e(i=1)$ \\
2 & 2 & $\beta_2$ & $\alpha_2$ & $e \mid (\beta_2,\alpha_2)$ & $y_2 \mid e(i=2)$ \\
3 & 2 & $\beta_3$ & $\alpha_2$ & $e \mid (\beta_3,\alpha_2)$ & $y_3 \mid e(i=3)$ \\
4 & 3 & $\beta_4$ & $\alpha_3$ & $e \mid (\beta_4,\alpha_3)$ & $y_4 \mid e(i=4)$ \\
... & ... & ... & ... & ... & ... \\
i & $\omega$ & $\beta_i$ & $\alpha_{\omega}$ & $e \mid (\beta_i,\alpha_{\omega})$ & $y_i \mid e(i)$\\
\botrule
\end{tabular}
\end{minipage}
\end{center}
\end{table}
All the nodes within the same $\omega$-clique would be connected to each other. A number of other links would be drawn between pairs of $(i,j)$ nodes, with a probability that is inversely proportional to $\mid\beta_i - \beta_j\mid$.
\begin{equation}\label{Eq:InverProp}
        Pr.(i_{\beta_i} \leftrightarrow j_{\beta_j}) \propto \frac{1}{\mid \beta_i - \beta_j \mid}
    \end{equation}
and $k_i$ would be the sum of all the links connecting $i$ to other nodes (degree of $i$), without distinction between linked nodes sharing the same clique of $i$ or not.

In practice, for networks of large size, this method is computationally expensive, because it would require to compute $\mid\beta_i - \beta_j\mid$ for each pair of nodes: these operations will happen in an exponential time.

\begin{figure}[h]%
\centering
\includegraphics[width=0.9\textwidth]{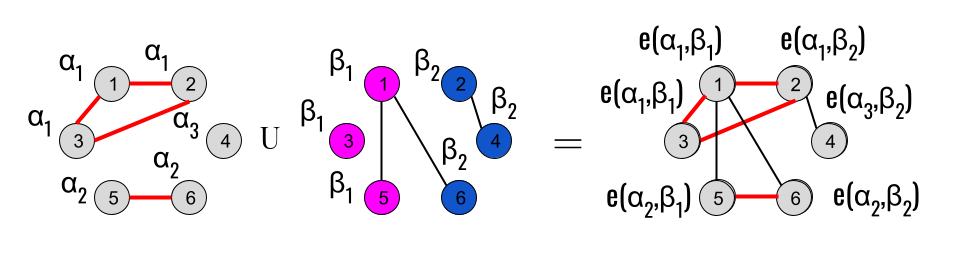}
\caption{The left graph is an ensemble of cliques. Each clique is associated to a $\alpha$ random value, shared among all the members of the clique. It can be noticed that a clique consists of only one node. Colours of nodes in the center graph represent membership of two different blocks. In the right graph, the edges of the center graph have been exported into the left graph, alongside the $\beta$ value associated to their block. $\alpha$ and $\beta$ will mix according to Eq. \ref{eq:p} into the risk score $e(i)$. }
\label{fig:cb}
\end{figure}

An efficient way to reduce the computation to a linear time is to adopt a stochastic blockmodel \citep{rohe_note_2018}. In the previous case, nodes would have been added individually to the pre-existing network of cliques. In this case, the set of edges from the cliques is engrafted with a set of edges generated from a stochastic blockmodel. For this reason, we call this model of network generation: `cliques-and-blocks'. The concept can be visualized in Figure \ref{fig:cb}.

Adopting cliques-and-block, minor adjustments occur to Eq. \ref{Eq:InverProp}. Each block would ideally represent a \textit{level of risk}, that is an ordinal category along the unit interval $[0,1]$. As a consequence, aforementioned distribution $D_\alpha = D_\beta = D$ can only be discrete since the numbers of blocks is finite. Assortativity can be parameterised through a mixing matrix $\textbf{B}$ that associates the probability that nodes in one block would link with nodes in another block.

\begin{figure}[h]%
\centering
\includegraphics[width=0.9\textwidth]{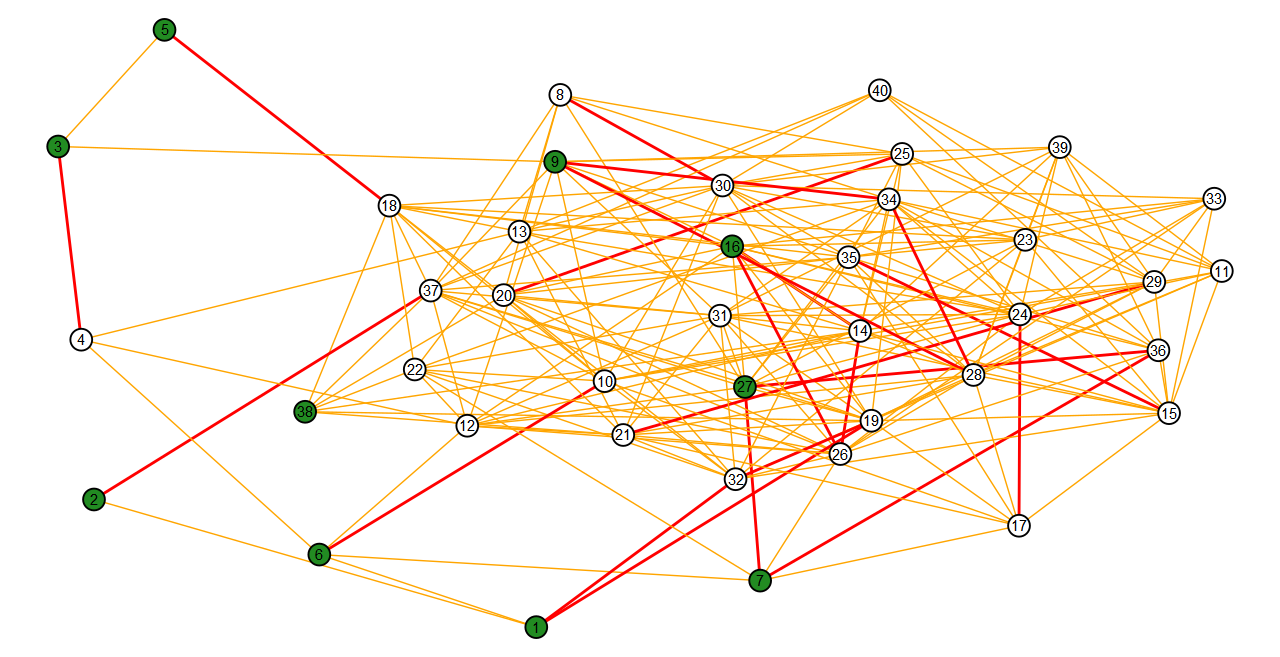}
\caption{Toy graph made of only 38 nodes. The dark green nodes are the smokers. Red edges are the cliques, while orange edges are imported from the stochastic blockmodel. In this graph, the parameter that regulates the isolation of smokers has been set very high to visualize how `cliques-and-block' can generate artificial isolation of a group without inducing modularity of in the network. In the simulation, the number of nodes can exceed $30.000$ and usually smokers are less isolated.}
\label{fig:toynet}
\end{figure}

In Fig. \ref{fig:toynet} it is represented a small network of 38 nodes that has been generated through the cliques-and-blocks simulation methodology. 11 of these ($28\%$) are smokers (in dark green). It can be noticed that many smokers, but not all of them, are isolated from the more dense area of the network. These dark green nodes are still quite well-connected with other dark green nodes.

In this simulation, operations are simulated with the help of softwares \texttt{igraph}, \texttt{tidyverse}, \texttt{tidygraph}, \texttt{Matrix}, and \texttt{fastRG}.

\subsubsection{Model Parameterisation}\label{Param}

Cliques are drawn with an internal size equal to:
\begin{equation}\label{eq:cliques}
\mathbb{E}(n_{\omega})\sim \mathcal{P}(1.2)+1
\end{equation}
The reference for the $\lambda$ parameter in Eq. \ref{eq:cliques} is the average number of members of households in Western Europe in 2020. 

$10$ levels of risk (discrete coefficients) are parameterised as $\beta$-blocks, ranging from $\beta = .05$ to $\beta = .95$. Given that this statistic is discrete and constrained, the probability mass function of density (PMF) for $D$ can be modeled after a $\mathcal{B}$inomial. However, if so, the variance of $e(i)$ would be lesser than the $\mathcal{B}$inomial model, since applying the formula in Eq. \ref{eq:p} it follows:
\begin{equation}\label{eq:variance}
\sigma([\mathcal{B}(9,p)] \cdot w + [\mathcal{B}(9,p)] \cdot [1-w]) \leq \sigma(\mathcal{B}(9,p)); \forall (p,w)
\end{equation}

\begin{figure}[h]%
\centering
\includegraphics[width=0.8\textwidth]{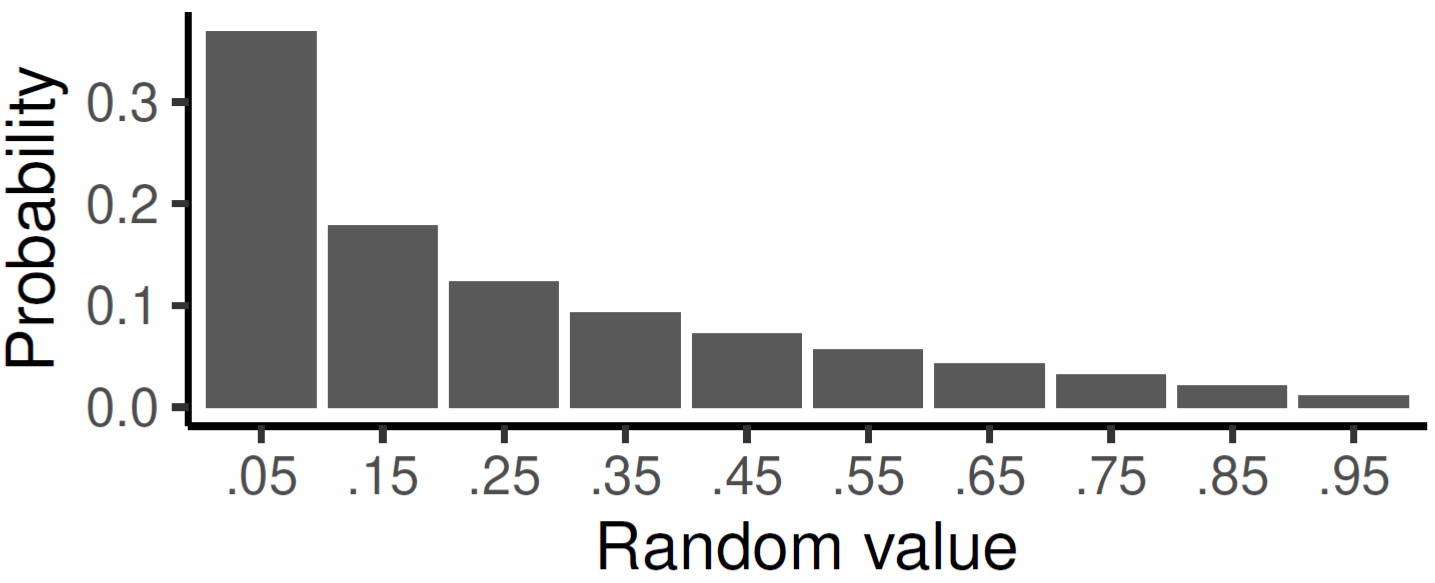}
\caption{Example of how probability is distributed across 10 ordered categories in an Overdispersed $\mathcal{B}$inomial model with parameter of centrality equal to $.275$ and parameter of dispersion $\vartheta = .3$. The 10 categories are the blocks of the cliques-and-block models, hence they are levels of risk of being a smoker. Higher levels of risks are always very rare in the simulation, but they are more likely when the parameter $p_D$ is higher.}\label{fig:PMF_TPois}
\end{figure}

\begin{figure}[h]%
\centering
\includegraphics[width=0.8\textwidth]{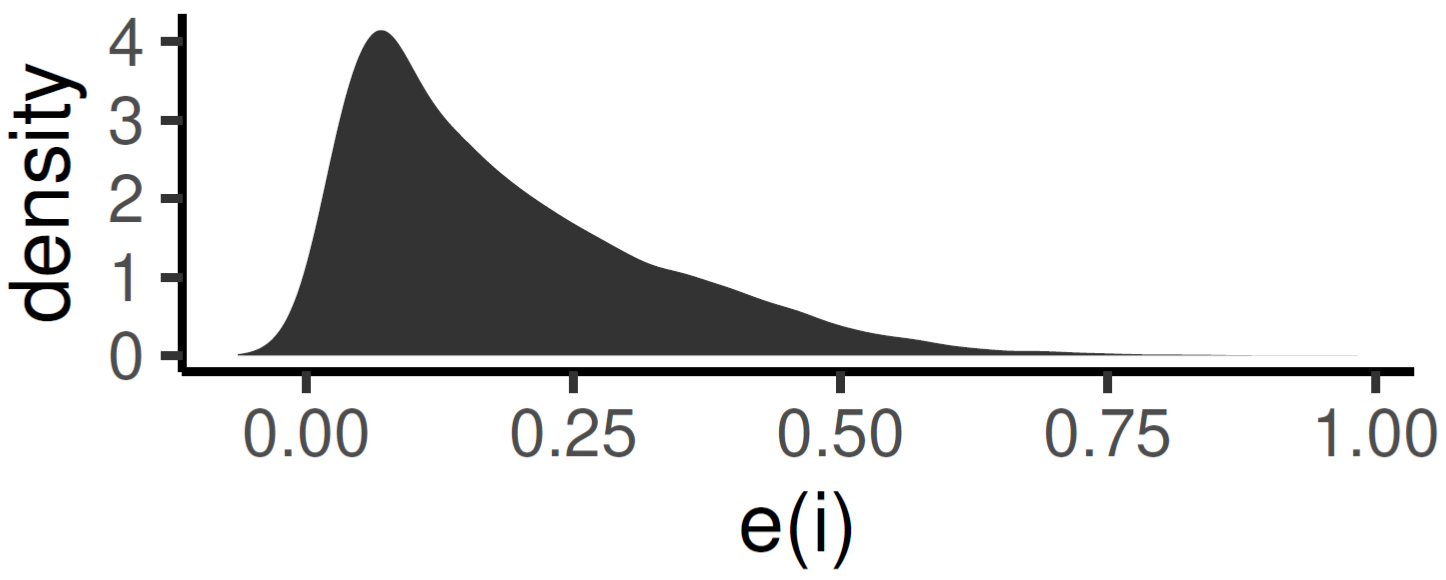}
\caption{Probability Density Function (PDF) of the $e(i)$ risk scores in a run with $19.413$ nodes, $w=.15$, $p_D = .21$. $e(i)$, and $\vartheta = .3$. $\alpha_i$ and $\beta_i$ are generated through Eq. \ref{eq:betabinom}, then they are mixed into the risk scores $e(i)$ (Eq. \ref{eq:p}). This example satisfies the principle enunciated in Eq. \ref{eq:betabinom2}, since $\bar{e}(i) = .185$ and $\sigma^2(e(i)) = 19.6$.}
\label{fig:e}
\end{figure}

To solve this issue, we adopt an Overdispersed $\mathcal{B}$inomial model \citep{prentice_binary_1986, moore_robust_1991}\footnote{We adopted the command \texttt{VGAM::rbetabinomial} in R.}:
\begin{equation*}
p_D \sim \mathcal{U}niform(.15,.30)
\end{equation*}
\begin{equation}\label{eq:betabinom}
D = Over\mathcal{B}in.(9,p_D,\vartheta_D) \cdot .1 +.05
\end{equation}
According to our calibrations, for the target $p_D$ in Eq. \ref{eq:betabinom}, a fixed overdispersion parameter $\vartheta_D = .3$ stabilizes the variance as:
\begin{equation}\label{eq:betabinom2}
\sigma^2(D) \sim \mathbb{E}(D)
\end{equation}
in most of the cases. The PMF for the average case of $p_D + .05 = .275, \vartheta_D = .3$ is provided in Fig. \ref{fig:PMF_TPois}. In Figure \ref{fig:e} is represented the distribution of $e(i)$ in a run of the algorithm.

Undirected edges are parameterised after a mixing matrix of $\beta$-blocks, that, for the case of undirected graph, is a symmetric square matrix $\textbf{B}: \pi_{D_{\beta}}$, where $\pi_{D_{\beta}}$ is the propensity that a $i$ node within block $\beta$ is connected to a $j$ node in a different block $\beta'$.

The algorithm assigns $\pi_{\beta,\beta'}$ following the PDF of the Normal distribution. Here it is documented for \texttt{R} language.

\begin{minipage}{\hsize}%
\lstset{frame=single,framexleftmargin=-1pt,framexrightmargin=-17pt,framesep=12pt,linewidth=0.98\textwidth,language=R}
\begin{lstlisting}
library(purrr)
library(Matrix)

map_dfc(
  seq(1,10,1),
  function(.x) {
    (dnorm(seq(1,10,1),.x,.x)
    )           }
        ) |>
        as.matrix() |>
  forceSymmetric(uplo = "U") |>
  as.matrix() -> B

B/sum(B) -> B
\end{lstlisting}
\end{minipage}

The core of the algorithm is that it associates the baseline $\pi_{\beta}$-propensity (for nodes $i$ from block $\beta$) to be linked to nodes $j$ from the same $\beta$-block to the normalised\footnote{It holds $\sum{\pi_{D_\beta}} = 1$, that is guaranteed through the imputation \texttt{B/sum(B) -> B}.} density of the probability to observe a random value $x: x = \mu$ from a $\mathcal{N}$ormal distribution that holds $(\beta+.05) \cdot 10$ both as location and standard deviation parameters\footnote{$Pr.$ indicates the mass of probability of a discrete value, $P$ indicates the density of probability of a point value in a continuous distribution}:
\begin{equation}\label{Eq:InverP}
    Pr.(i_{\beta} \leftrightarrow j_{\beta}) \propto P((\beta+.05) \cdot 10 \in \mathcal{N}(\mu = \sigma = (\beta+.05) \cdot 10)
\end{equation}
Likewise, each deviation of $\pi$ for propensity of nodes connecting with nodes from another block ($i_{\beta} \leftrightarrow j_{\beta'}$) is modeled after the normalised density of probability for the deviation from locations in $\mathcal{N}(\beta,\beta)$:
\begin{equation}\label{Eq:InverP2}
Pr.(i_{\beta} \leftrightarrow j_{\beta'}) \propto P((\beta'+.05) \cdot 10 \in \mathcal{N}(\mu = \sigma = (\beta+.05) \cdot 10)
\end{equation}
As a model for a baseline mixing matrix, this satisfies two desired axioms that are discussed extensively in the Section \ref{Theo}:
\begin{enumerate}
    \item From Eq. \ref{Eq:InverP} and Eq. \ref{Eq:InverP2}, it follows that each block always has a higher propensity to link within itself than any other block, hence the network will be smoking-assortative.
    \item Lower $\beta$-blocks are more connected than higher $\beta$-blocks, hence non-smokers have higher $k_i$-degree than smokers.
\end{enumerate}

This algorithmic approach to network generation, does not  raise modularity of smokers (i.e., divisibility of the graph in clusters of smokers vs. non-smokers). As a consequence, since there are more non-smokers than smokers (see the distribution of parameter $p_D$ in Eq. \ref{eq:betabinom}), the networks will also show degree-assortativity.

Memberships of a $i$-node to a $\omega$-clique and to a $\beta$-block are mutually independent. This assumption is not discussed in Section \ref{Theo}, and it could be relaxed in more refined parameterisations (discussed in Section \ref{Disc}).

\begin{figure}[h]
\label{fig:B2}
\centering
\includegraphics[width=0.9\textwidth]{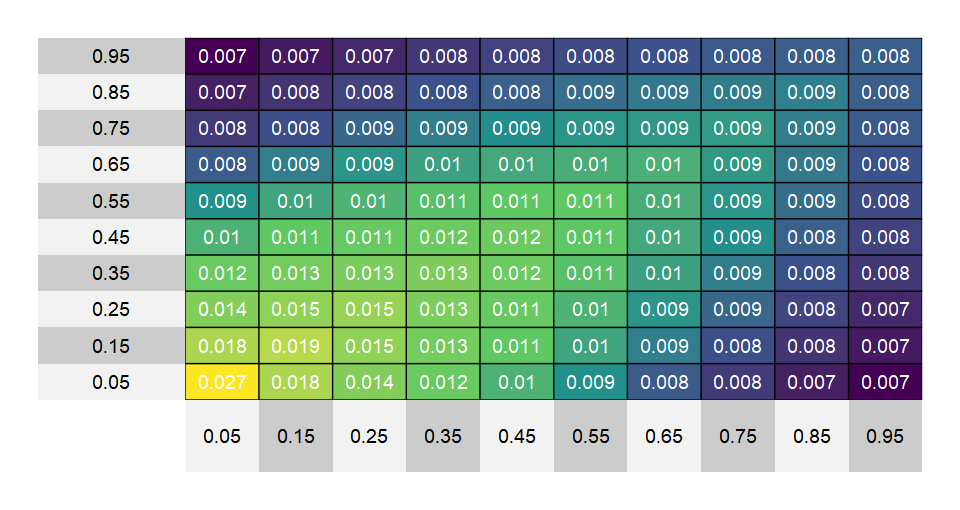}
\caption{This mixing matrix has been generated applying the algorithm and then exponentiating each $\pi_{\beta,\beta'}$ element of the baseline matrix \texttt{B} to $\gamma = .5$. One can notice that diagonal of the matrix follows a geometric progression such that the difference between $\pi_{\beta,\beta}$ and $\pi_{\beta+.05,\beta+.05}$ is always the half of the difference between the difference between $\pi_{\beta-.05,\beta-.05}$ and $\pi_{\beta, \beta }$. This propriety hold $\forall \gamma$, $\gamma$ only regulates the statistical distances between the diagonal and the other elements of the matrix, and indirectly the homophily in the network.}
\end{figure}

The axioms can be both visually checked in Fig. \ref{fig:B2}. Dispersion of the $\pi_{D_{\beta}}$ can be controlled through a re-parameterisation with a $\gamma$ exponent\footnote{Another technique to alter the deviations in $\textbf{B}$ is through matrix exponentiation:
\begin{equation*}\label{eq:gamma2}
   \textbf{B}: (\pi_{D_{\beta} \mid \gamma}) = \frac{\textbf{B}^{\gamma}}{\sum{\pi_{\textbf{B}^{\gamma}}}}
\end{equation*}
instead of exponentiation of the element.

This is not suited for this work in particular because, for matrix exponentiation $\exists \gamma$ such that axioms $1.$ and $2.$ would be violated.

}:
\begin{equation}\label{eq:gamma}
   \pi_{D_{\beta} \mid \gamma} = \frac{\pi_{\textbf{B}}^{\gamma}}{\sum{\pi_{\textbf{B}}^{\gamma}}}
\end{equation}

For any practical evaluation of the role of $\gamma$ in the network generation model, it can be said that the higher the $\gamma$, the higher will be attribute-homophily in the network, hence the degree-homophily, too.

The stochastic blockmodel generation is run through software \texttt{fastRG}. \texttt{fastRG} will generate a composite population of links and the distribution of the degree of the links of the blockmodel $k_{\beta}$ will follow a $\mathcal{P}$oisson model.

As anticipated in Table \ref{tab:ei}, once the network has been generated, to each $i$-node is assigned the binary $y_i$ with:

\begin{equation}\label{eq:y}
 Pr.(y_i = 1) = Bernoulli(e_i)
\end{equation}

\subsection{Recruitment}\label{Recru}

In order to evaluate the alternative outcome of switching from gold standard to HPSSD, $8,000$ $v$-runs are generated following the model in Section \ref{PopModel}. $y$ denotes the quota of nodes with a positive outcome in Eq. \ref{eq:y}, so it holds $y_v = \bar{y}_{i \in v}$.

For each run, a universal parameter of attrition $r_v$ is randomly set. Each node at $v$ has an individual parameter of attrition $r_i$:
\begin{equation*}
p_r = r_v + (\frac{e_i - \bar{e}_{i})}{10}) \cdot .25    
\end{equation*}
\begin{equation} \label{eq:ri2}
    r_i \sim \frac{\mathcal{B}(100,p_r)}{100}
\end{equation}

$\lfloor \frac{1,000}{(1-r_v)} \rfloor$ nodes are uniformly random drawn from the population for each $v$-run. Each of these nodes is be pooled in or out into a benchmark sample with a probability equal to $r_i$. This sample is referred as $C_{\bigstar}$ or `the golden sample' for $v$. The average of the golden sample $\bar{y}_{\bigstar}$ is the gold standard estimate.

From here, for each run, four parallel processes of recruitment are initiated from the golden sample, each representing a different scenario of HPSSD within the random parameterisation featured in $v$. These scenarios are alternative ``what if?", so they could be evaluated comparatively as proper alternative outcomes for a casual analysis of the effect of switching from a benchmark to HPSSD.

\subsubsection{Scenario I: Full sample, low dispersion}

In this scenario, the golden sample and the stage 0 for HPSSD are the same set  $C_{\bigstar} = C_{t_0}$.

Each seed recruits $m_i$ new $j$-nodes into the next stage of respondents $t_1$. The number of recruited by $i$ ($m_i$) follows a $\mathcal{P}$oisson model ($\lambda := .5$), and if it exceeds $k_i$, it is floored into $k$. Each of the $j$ recruited nodes is pooled into the stage $t_1$ with a probability equal to $1 - r_j$. Also, if $j \in n_{<t}$ already, then it is excluded from $n_{t}$. This is iterated through $t$ stages until $n_t = 0$.

As a consequence of $\lambda := .5$, it holds:
\begin{equation}\label{eq:ri}
    \mathbb{E}(n_t) \sim \frac{(1-r) \cdot n_{(t-1)}}{2} 
\end{equation}
so it follows
\begin{equation}\label{eq:rv}
    r_v = 0 \Rightarrow \mathbb{E}(n_{t>0}) \sim \mathbb{E}(n_0)
\end{equation}
so we expect the sample size to be roughly the double of the seeds, given no attrition.

In this scenario, the union $\bigcup i \in t$ constitutes the hybrid sample $C_{I}$ of the I scenario. $C_{I}$ allows to evaluate performance of HPSSD assuming parity of operational cost between $C_{\bigstar}$ and $C_{I}$.

\subsubsection{Scenario II: Full sample, high dispersion}

This scenario is identical to I, with the difference that for $m_i$, instead of a $\mathcal{P}$oisson model is adopted a (shifted) $\mathcal{Y}$ule model\citep{huillet_new_2020}. $\mathcal{Y}$ule is a mono-parametric discrete distribution within the family of Power Laws\footnote{The unshifted model has been formalised by Yule. It is also known as Yule-Simon because Herbert Simon (\citeyear{simon_class_1955}) linked the model to the Zipf's Law and the Principle of Least Effort.}. The PMF for a shifted $\mathcal{Y}$ule distribution is:

\begin{equation}\label{eq:PMFyule}
    f(k \in \mathbb{N}^0 \mid \lambda) = \frac{\lambda \cdot (\lambda! \cdot k!)}{(\lambda + k + 1)!}
\end{equation}
and it holds
\begin{equation}\label{eq:Eyule}
    \mathbb{E}(k \mid \lambda) = \frac{\lambda}{\lambda - 1} -1 
\end{equation}

From Eq. \ref{eq:Eyule} it follows that a shifted Yule process converges to finite values $\forall \lambda > 2$, and this convergence is strong for $\lambda > 3$. Therefore, to preserve the principle of Eq. \ref{eq:rv}, from Eq. \ref{eq:Eyule} it follows that $\lambda_{\mathcal{Y}ule} := 3$.

In this scenario II, few seeds would be responsible for the majority of the snowball component of the hybrid sample $C_{II}$, hence the mention of ``high dispersion".

\subsubsection{Scenario III: Half sample, low dispersion}

This scenario is identical to I, with the only difference that in this case $C_{t_0}$ is half of the size of $C_{\bigstar}$, by discarding half of the members of $C_{\bigstar}$.

\subsubsection{Scenario IV: Half sample, high dispersion}

This scenario combines both the features of Scenario II and Scenario III.

\subsection{Evaluation Strategy}

In Section \ref{Recru} all of the four scenarios start with a common random sample $C_{\bigstar}$ that is also the benchmark sample for that run. The process that samples $C_{scenario}$ from  $C_{\bigstar}$ can be seen, conceptually, as a causal intervention, since it preserves random-generated features inherent to that run as a form of \textit{ceteris paribus}. This specific design allows to compute evaluative statistics on differences within the same run. These differences can be interpreted as specific performances of the HPSSD samples, given $C_{\bigstar}$ as a benchmark.

Differences in absolute errors between the two samples measure the performance of HPSSD:
\begin{equation} \label{Eq:perform}
\Delta = \frac{\mid y - \hat{y}_0 \mid - \mid y - \hat{y}_1 \mid}{y} 
\end{equation}
where $\hat{y}_1$ represents the estimate of $y$ according to the benchmark design, and $\hat{y}_0$ according to the alternative proposal, i.e. HPSSD in a scenario. That means that if $\Delta$ is $>1$, then to adopt the alternative would have been beneficial in that run, because it would minimize the expected margin of error. However, the difference still depends by the prevalence in the population $y$, hence in Eq. \ref{Eq:perform} it is divided to $y$. 
This operation allows to employ a summary statistic of $\Delta$ across the runs as an estimate of the expected improvement in the margin of error after switching to the alternative. In other words, given that the runs are all generated independently, $\bar{\Delta}$ works as an estimate of the net benefit, expressed as a rate of increase of performance expected after switching into HPSSD. A negative value would indicate a worse performance.

A second proposal for performance evaluation is non parametric:

\begin{equation} \label{Eq:nonpara}
\zeta = \frac{n(\mid y - \hat{y}_0 \mid > \mid y - \hat{y}_1\mid)}{N} 
\end{equation}

Eq. \ref{Eq:nonpara} does not estimate the net benefit of switching from one design to another, but the expected relative frequency that the alternative will outperform the standard. This statistic is easier to be interpreted and, combined to $\bar{\Delta}$, it should provide the whole picture on the results of the simulation.

Determinants of the errors are \textit{variance} and \textit{bias}. Being random sampling unbiased, error of gold standard is always due to inherent variance given the sample size. HPSSD has always a sample size that is higher of its random stage 0, so the expected error component due variance should be inferior. However HPSSD is also biased due to homophily in the population. In other words, when switching to HPSSD there is a trade-off between a reduction in variance and an increase of bias.

The Design Effect is the analytical statistic to evaluate the expected reduction of variance in a sample estimator that is alternative to gold standard:
\begin{equation} \label{Eq:DE}
DE = \frac{\sigma^2(\hat{\theta_1})}{\sigma^2(\hat{\theta_0})}
\end{equation}
where $\hat{\theta_0}$ represents the random sampling estimator and $\hat{\theta_1}$ the alternative proposal. With minor adaptations, from Eq. \ref{Eq:DE} it can derived a statistic for the evaluation of the rate of reduction in random error after switching to the alternative:
\begin{equation} \label{Eq:psi}
\psi = 1- \frac{s^2(y - \hat{y}_1)}{s^2(y - \hat{y}_0)}
\end{equation}

Estimation of bias is straightforward, given independent runs. Since bias is nothing more than the location of the errors of a design,
\begin{equation} \label{Eq:bias}
bias = 	\langle y - \hat{y} \rangle
\end{equation}
it follows
\begin{equation} \label{Eq:biashat}
\hat{bias} = avg(y - \hat{y})
\end{equation}

\section{Results}\label{Resu}

$8,000$ $v$-runs have been randomly generated with the cliques-and-block engrafting model described in Section \ref{PopModel}.
A run is a population of nodes connected in a graph. A $y$ quota of this population is made of target units with a ('smoker'). $y$ is also the estimand of the sampling procedures that are compared in this study. General features of the populations are:
\begin{itemize}
    \item nodes are fully connected in very small clusters of few nodes (cliques) and sparsely connected with nodes outside their clique.
    \item the network is both $y$-assortative and degree-assortative
    \item target nodes ($i \mid y = 1$) are more isolated than non target nodes ($i \mid y = 0$)
    \item all nodes have a propensity to non-response (individual attrition) to a survey and this individual attrition is slightly higher in target nodes.
\end{itemize}
To achieve variability in the intensity of these features, each run is distinct from the others through the variability of 6 parameters (see, Table \ref{tab:params}).

\begin{table}[h]
\begin{center}
\begin{minipage}{240pt}
\caption{Parameters of the network population.}
\label{tab:params}%
\begin{tabular}{@{}llrl@{}}
\toprule
Symbol & Concept & Range & Documented\\
\midrule
$y$ & Target Quota & $[.15,.35]$ & Eq. \ref{eq:y} \\
$\omega$ & Number of cliques & $[5.000,15.000]$ & Eq. \ref{eq:cliques} \\
$\langle k \rangle$  & Ego-nets size & $\sim [5,25]$  & Sec. \ref{Param} \\
w & Familism & $[.1,.5]$ & Eq. \ref{eq:p} \\
$\gamma$ & Homophily & $[.2,.8]$ & Eq. \ref{eq:gamma}  \\
r & Attrition & $[0,.5]$ & Eq. \ref{eq:ri2}\\
\botrule
\end{tabular}
\end{minipage}
\end{center}
\end{table}

\begin{figure}[h]%
\centering
\includegraphics[width=0.9\textwidth]{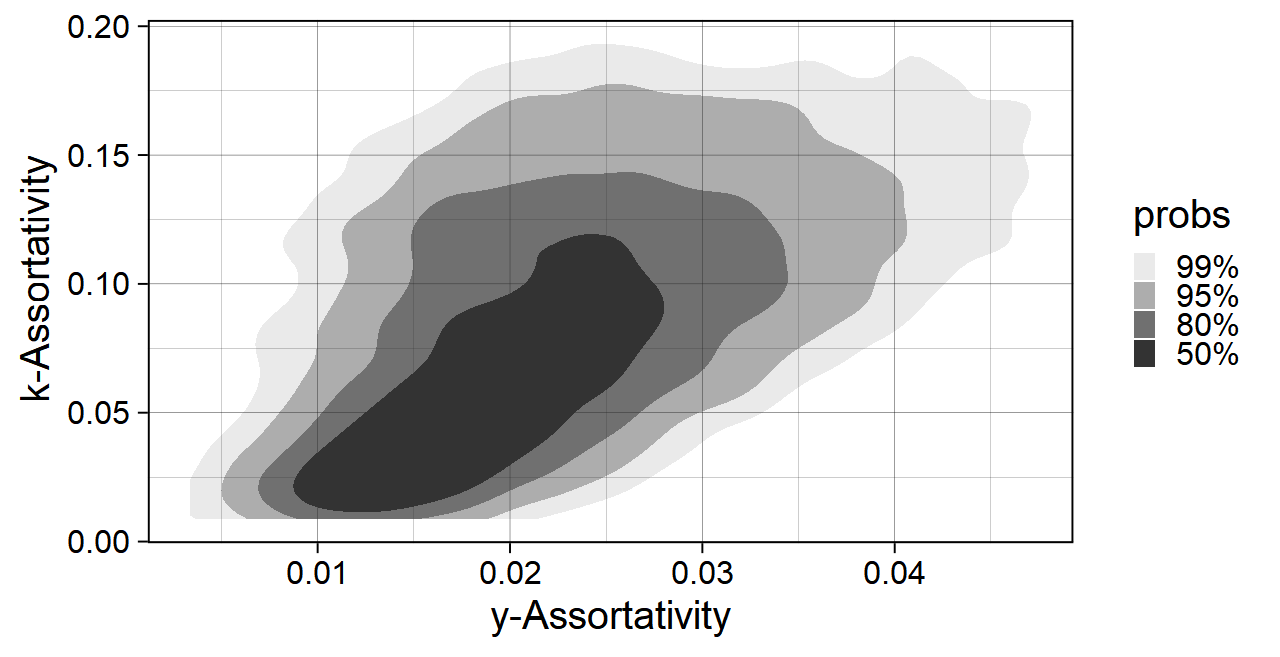}
\caption{Joint density of frequency of $\phi_y$ (x-axis) and $\phi_k$ (y-axis), across the $8,000$ runs. Colours represent how much of the sample is frequent in the colored area. Half of the density is in the dark area.}\label{fig:phis}
\end{figure}

In particular, through Eq. \ref{eq:gamma}, $\gamma$ determines jointly the levels of degree-homophily $\phi_k$ and $y$-homophily $\phi_y$. Across the runs, $\phi_k$ is both higher and more variable than $\phi_y$ (Fig. \ref{fig:phis}).

Indeed, across the variety of parameters under analysis in this study, the homophily parameter is the second most predictive of the absolute error of HPSSD across the different scenarios, only behind the value of the estimate of $y$ in $t_0$ (Table \ref{tab:resu1}). 

\begin{table}[h]
\begin{center}
\begin{minipage}{315pt}
\caption{Standardized bivariate regressions on the absolute error in each scenario.}\label{tab:resu1}%
\begin{tabular}{@{}llrrrr@{}}
\toprule
& & \multicolumn{4}{c}{\textit{Scenarios}}\\
\cline{3-6} 
Regressor & Concept & I & II & III & IV\\
\midrule
$\hat{y}_0$ & Stage 0& 0.373$\pm.01$ & 0.379$\pm.01$ & 0.237$\pm.011$ & 0.230$\pm.011$\\
$\gamma$ & Homophily & 0.207$\pm.01$ & 0.211$\pm.01$ & 0.146$\pm.011$ & 0.137$\pm.011$\\
w & Familism & -0.112$\pm.01$ & -0.108$\pm.01$ & -0.090$\pm.011$ & -0.093$\pm.011$\\
r & Attrition\ & -0.093$\pm.01$ & -0.085$\pm.01$ & -0.045$\pm.011$ & -0.045$\pm.011$\\
$\langle k \rangle$ & Ego-nets size & 0.082$\pm.01$ & 0.086$\pm.01$ & 0.051$\pm.011$ & 0.045$\pm.011$ \\
y & Target Quota & 0.053$\pm.01$ & 0.065$\pm.01$ & 0.054$\pm.011$ & 0.075$\pm.011$\\
N & Pop. Size & 0.030$\pm.011$ & 0.035$\pm.011$ & 0.014$\pm.011$ & 0.016$\pm.011$\\
\botrule
\end{tabular}
\footnotetext{Values in the columns are the coefficients of $\mid y - \hat{y}_{scenario} \mid \; \sim Regressor$, plus the standard error of the coefficients for each scenario. These are computed on the Monte Carlo sample of the simulation, made of $8,000$ runs. Mathematically, these values will fall very close to the values of linear correlation coefficients in those cases where intercepts are $0$ by design. All the coefficients are associated to \textit{p}-value very close to $0$.}
\end{minipage}
\end{center}
\end{table}

In Table \ref{tab:resu1}, attrition shows a negative coefficient. That would imply that more attrition leads into lower margin of error. It requires an explanation: the model links $n_\bigstar$ to the expected attrition (see, Section \ref{Recru}) in order to fix $n_\bigstar \sim 1,000$. Higher attrition would still impact $n_{scenario}$ reducing the snowball component, hence increasing sample variance. However, across the majority of the runs, the snowball quota has a negative impact on the performance of the estimator $\hat{y}$ for sample scenarios I and II (see, in Table \ref{tab:resu2}, row ``ALL").

\begin{table}[h]
\begin{center}
\begin{minipage}{190pt}
\caption{Average improvement in the margin of error in scenarios}\label{tab:resu2}%
\begin{tabular}{@{}lrrrr@{}}
\toprule
& \multicolumn{4}{c}{\textit{Scenarios}}\\
\cline{2-5} 
Homophily & I & II & III & IV\\
\midrule
Low & 0.25\% & 0.25\% & 0.76\% & 0.75\% \\
Mid-Low & -0.37\% & -0.36\% & 0.5\% & 0.44\% \\
Mid-High & -0.87\% & -0.8\% & -0.3\% & -0.17\% \\
High & -1.73\% & -1.73\% & -0.92\% & -0.92\% \\
ALL & -0.68\% & -0.66\% & 0.01\% & 0.03\% \\
\botrule
\end{tabular}
\footnotetext{Values are $\bar{\Delta} \cdot 100$. Scenarios are crossed scenarios with four quartiles of the distribution of $\gamma$ across the $8,000$ runs of the simulation. A positive $\Delta$ indicates that switching to the HPSSD in that run would decrease the absolute error in the estimate of $y$.}
\end{minipage}
\end{center}
\end{table}

As a consequence, reducing the snowball quota through higher attrition, in many cases, improves the estimate. This hypothesis has been checked through a multivariate regression:
\begin{equation} \label{Eq:rcheck}
\Delta \sim a + b_1(n - n_0) + b_2(\gamma) + b_3(r)
\end{equation}
For the I scenario, the $b_3$ coefficient for $r$ is positive ($.15 \pm .034$), but still inferior to $b_2$ coefficient for gamma ($.188 \pm .01$) and $b_1$ coefficient for $n - n_0$ ($.247 \pm	.034$), and similar results hold for the other scenarios.

Across the scenarios, when not controlled \textit{per gamma}, $\bar{\Delta}$ is negative or trivially positive (see ``ALL" in Table \ref{tab:resu2}). However, for low level of homophily the margin of error is reduced, even if for abysmal quantities (0.25\% is $.0025$ of the error, less than 1\%). These very small values lead to conclude that the impact of the snowball component of HPSDD is very little, too. At the same time, the result that $\bar{\Delta}$ is higher for the III and IV scenarios (when $y_0$ is computed on half of the golden sample) suggests a conclusion: for null or low levels of homophily HPSSD is still a viable design to reduce the margin of error if the research has financial hardship to reach a representative sample. This could be the case for many small scientific projects that cannot fall into the lens subject general national surveys.

Researchers should always primarily aim to reach representative sample size with a randomised design, and only is that is not wholly possible, then resort to augment it with a snowball component. HPSSD cannot be justified by the unexpensive increase in sample size alone, because the $\Delta$ of the absolute errors is very sensitive to variability in the homophily parameter (see Fig. \ref{fig:phis}). The Table \ref{tab:zetas} of $\zeta$ support this suggestion.

\begin{table}[h]
\begin{center}
\begin{minipage}{130pt}
\caption{Relative frequency of runs where gold standard has a higher absolute error than HPSSD, given the scenario.}\label{tab:zetas}%
\begin{tabular}{@{}lrrrr@{}}
\toprule
& \multicolumn{4}{c}{\textit{Scenarios}}\\
\cline{2-5} 
Homophily & I & II & III & IV\\
\midrule
Low & .52 & .52 & .55 & .55\\
Mid-Low & .44 & .44 & .54 & .54\\
Mid-High & .38 & .39 & .45 & .47\\
High & .29 & .31 & .40 & .40\\
\botrule
\end{tabular}
\end{minipage}
\end{center}
\end{table}

\subsection{De-biased HPSSD estimates}

Estimation of $y$ through the sample mean of HPSSD is biased, however when homophily is lower and the bias is trivial, it holds a lesser margin of error than the costwise random standard, because the random component of the error is reduced through the increase in sample size (Table \ref{tab:psibias}).

\begin{table}[h]
\begin{center}
\begin{minipage}{140pt}
\caption{Reduction of variance, and estimate of bias, across the scenarios}\label{tab:psibias}%
\begin{tabular}{@{}lrrrr@{}}
\toprule
& \multicolumn{4}{c}{\textit{Scenarios}}\\
\cline{2-5} 
& I & II & III & IV\\
\midrule
$\psi$ & 0.28\% & 0.26\% & 0.31\% & 0.30\%\\
$\hat{bias}$ & -.01 & -.01 & -.01 & -.01\\
\botrule
\end{tabular}
\end{minipage}
\end{center}
\end{table}

Removing the average bias $-.01$ from $\bar{y}$, for all levels of homophily HPSSD performs better than its own random component $C_0$ (Table \ref{tab:unbiased})

\begin{table}[h]
\begin{center}
\begin{minipage}{135pt}
\caption{Relative frequency of runs where gold standard has a higher absolute error than de-biased HPSSD, given the scenario.}\label{tab:unbiased}%
\begin{tabular}{@{}lrrrr@{}}
\toprule
& \multicolumn{4}{c}{\textit{Scenarios}}\\
\cline{2-5} 
Homophily & I & II & III & IV\\
\midrule
Low & .58 & .56 & .58 & .58\\
Mid-Low & .60 & .60 & .61 & .62\\
Mid-High & .60 & .60 & .59 & .60\\
High & .55 & .55 & .57 & .56\\
\botrule
\end{tabular}
\end{minipage}
\end{center}
\end{table}

Even if the addition of $.01$ to the estimate does not equate with an exact correction for the unbiased estimation\footnote{Since the correction should account for the individual features that can be inferred on the run under observation, and not only for a global statistic}, now is less risky to adopt HPSSD because $\zeta > .5$ in the de-biased estimates (see Table \ref{tab:unbiased}))

However, without a solid prior knowledge for the levels of homophily in the real population to sample, the proposed de-biasing correction would only show that there is potential for true unbiased estimators of HPSSD that would perform better than the costwise random alternatives.

\section{Discussion}\label{Disc}

This study has been conducted having in mind specific heuristics of applied Statistics. This would explain why $\lfloor \frac{1,000}{(1-r_v)} \rfloor$ is been elicited as the representative sample size for $C_\bigstar$, as if the practitioners have already an exact prior of the attrition - that is often the case for rigorous population studies. Researchers would know that a random sample of $\sim 1,000$ can infer binary features within virtually any human population with a margin of error inferior to 3\% and a Confidence Level of 95\%\footnote{This would also explain why there was no need for networks with more than $35.000$ nodes, as the effect of $N$ on $\hat{y}$ and $\Delta$ is abysmal (see Table \ref{tab:resu1}).}.

Addition of $.01$ stochastically improves the performance of HPSSD for all the tested levels of homophily in Table \ref{tab:unbiased}. Should this correction be regarded as a heuristic to improve estimates in absence of an analytically validated estimator? We suppose that our study only shows that there is potential for analytical proposal for an unbiased HPSSD estimator, and more research is needed before drawing conclusions on this subject.

There are two major issues in the analytical quest for a HPSSD: the first is that even if homophily has a relevant role in the bias of HPSSD, the model generating the population still has 6 primary hyperparameters (Table \ref{tab:params}). It is hard to validate an unbiased estimator through all the sources of variance of the model. At the same time, results show that homophily must be treated before the other parameters. In this sense, the lack of relevant differences between scenarios with low and high dispersion of recruitments is helpful, because it means that the unbiased estimator could be potentially agnostic in regards of the distribution of recruited \textit{per} respondent.

Vacca et al. (\citeyear{vacca_cross-classified_2019}) and Audemard (\citeyear{audemard_objectifying_2020}) are noteworthy for linking the inference in chained observations to inference in hierarchical data \citep{gelman_data_2007}. Actually, the general problem in the research of a unbiased estimator for snowball sampling is that at the current stage is very hard to have a prior knowledge on the assortativity in the network and/or in the forests (Fig. \ref{fig:forest}) within the sample space. 

Analogous issues would impact the adoption of the adjusted Volz-Heckathorn estimator \citep{volz_probability_2008}, that is the standard for RDS:
\begin{equation} \label{Eq:volz}
\hat{y} = \bar{y} \cdot  \genfrac{}{}{1pt}{}{\bar{k}}{(\bar{k} \mid y=1)}
\end{equation}
with the difference that while in RDS $k$ represents an estimate of the ego network size of the respondent unit $i$, in HPSSD $k$ is an observed value that is drawn within the limited small portion of the ego network of $i$. Here the limit is that in HPSSD, $k$ does not depend entirely on $i$, because in real cases $j$ could refuse-by-default to be recruited by $i$; that is a corollary on the main argument of Crawford et al (\citeyear{crawford_identification_2018}) on the modeling of preferential recruitment as a separated sub-process within RDS.

We spouse the idea that link-tracing technology, even as something separated from analytical inference, should be more adopted in population studies. In facts, the idea that a snowball sample in $t$-stages can be represented through a multilevel model stem from the fact that each unit belongs to a specific Galton-Watson tree. Especially for data not showing overdispersion of $k$ across the trees, the difference between the variance of the attribute \textit{between} the trees and the variance of the attribute \textit{within} the trees should be indicative of the presence of homophily in the networks. More in general, the benefit of metadata collection in sampling design has been historically underrated. For example, Liu and Stainbeck (\citeyear{liu_interviewer_2013}) found that gender and ethnicity of the interviewer is significantly correlated with gender and ethnicity of the respondents in the US General Social Survey. The assignment of interviewers is independent of the features of the drawn statistical units (e.g. households, phone numbers, etc.). Hence, excluding bad faith of the interviewers, this correlation can be possible only because the attrition of the potential respondents is influenced by the characteristics of the interviewer. This is an example that shows how the practice of tracking who is the interviewer has a direct impact in expanding the theory behind sampling designs for population studies.

This observation opens the discussion regarding two assumptions that constitutes limitations in the models of this present study:

\begin{enumerate}
    \item The assumption that $\alpha$ of the cliques and $\beta$ of the blocks are mutually independent implies that, in a human population, the propensity of the target variable that is derived by facts happening outside the household is uncorrelated with the characteristics that are shared within the household. In reality, this assumption would not always hold. For example, it is possible that poor families live in the same neighborhoods. Assuming that poverty is a driver for smoking \citep{giordano_impact_2011}, the likelihood to recruit a $j$-smoker from a $i$-poor would not only be influenced by $j$ and $i$ living in the same neighborhood, but also by the likelihood that the $j$ lives in a household where someone else smokes, an occurrence more likely in poor neighborhoods.
    The model does not assume influence effects outside households, but this is a very open controversy in science \citep{aral_distinguishing_2009, shalizi_homophily_2011}.
    \item The model has an assumption which considerably simplifies the computation: the network has not the necessity to represent those ties having a zero probability to be recruited. These connections are just not represented. In practice, it involves the choice of how to model the \texttt{k} parameter in \texttt{fastRG::sbm()} ($k_\beta$ in Section \ref{Param}) that, summed to the size of the clique of $i$, would determine $k_i$. In this study, $\langle k \rangle$ is modeled to not exceed $25$ (see Table \ref{tab:params}). This range is modeled following the study in human social networks of Robin Dunbar (\citeyear{dunbar_social_1998}), according to whom the number of very strong social ties is limited in humans and it varies between $5$ and $25$. This ego network represents family members (the clique) or close friends (the other edges). Dunbar's theory is paradigmatic in social networks and has been validated multiple times under different research frameworks \citep{goncalves_modeling_2011, west_relating_2020, dunbar_structure_2015}. However, it has also been disproved under both empirical \citep{mccarty_comparing_2005} and methodological \citep{lindenfors_dunbars_2021} arguments, too.
    The simplification that helps computation regards the fact that the $m_i$ recruited among the $j_i$-nodes in the ego-network of $i$ are drawn with a uniform probability. We do not think that this holds exactly, but that it holds stochastically, in the sense that, for example, within the specific ego-network of close peers, some people could be slightly more prone to recruit members of the family, others could be slightly more prone to recruit colleagues or friends, etc. So we assumed that overall the distribution of the probabilities should be uniform to represent equal levels of social proximity between $i$ and each $j_i$. This assumption is paradigmatic across the literature on RDS, even if Crawford et al. (\citeyear{crawford_identification_2018}) suggests that given the shortage of empirical validation, this could not be the case, at least on a theoretical level. Indeed, alternative models would involve a relation between $|e_i - e_j|$ and $Pr.(i \leftrightarrow j)$, but it would incur into expensive computations, as mentioned for Eq. \ref{Eq:InverProp}. In this sense technical developments for speeding up the computation are strongly suggested.
\end{enumerate}

There is a third argument worth of discussion that could be seen both as a limitation and a strength of the results, that is the principle encapsulated in Eq. \ref{eq:rv}: the snowball component is not expected to be much larger than the random component. In practice, even if snowball branching process should converge to a finite size, we expected much more variability in the relation between the quotas of random and snowball component of HPSSD. In the documented case of Etter and Perneger (\citeyear{etter_snowball_2000}) they are actually more or less of the same size, even if attrition was higher than zero, but this was reached under a peculiar design not implemented in HPSSD. To our knowledge, the premises of HPSSD such that the random component of the sampling should be semi-representative of the populations are not met in previous empirical studies. In absence of further evidences on the expected ratio between components, we believe that our choice has been conservative.

\section*{Acknowledgments}

We thank Prof. Harry Crane, Prof. Karl Rohe and Dr. Alexander Hayes for the time spent commenting ideas behind this study. The punctual insights that we received from Dr. Thomas L. Petersen, author of \texttt{tidygraph}, were precious.


\end{document}